\documentclass[twocolumn,showpacs,preprintnumbers,amsmath,amssymb,superscriptaddress,pra]{revtex4}
\usepackage[latin1]{inputenc}
\usepackage{dcolumn}
\usepackage{bm}
\usepackage{graphicx}
\usepackage{multirow}

\def\bbm[#1]{\mbox{\boldmath $#1$}}
\newcommand{\ket}[1]{\displaystyle{|#1\rangle}}
\newcommand{\bra}[1]{\displaystyle{\langle #1|}}
\newcommand{\sumpk}{\sum_p\int_0^{+\infty}dk_z\int d^2\mathbf{k}}

\newcommand{\apk}{a_p(\mathbf{k},k_z)}
\newcommand{\acpk}{a^\dag_p(\mathbf{k},k_z)}

\newcommand{\rv}{\mathbf{r}}

\begin{document}

\title{Dynamical aspects of atom interferometry in an\\optical lattice in proximity of a surface}
\author{Sophie Pelisson}\email{sophie.pelisson@obspm.fr}\affiliation{LNE-SYRTE, Observatoire de Paris, CNRS UMR8630,
UPMC\\61 avenue de l'Observatoire, 75014 Paris, France}
\author{Riccardo Messina}\affiliation{LNE-SYRTE, Observatoire de Paris, CNRS UMR8630,
UPMC\\61 avenue de l'Observatoire, 75014 Paris, France}
\affiliation{Laboratoire Charles Fabry, Institut d'Optique, CNRS, Universit\'e Paris-Sud,
Campus Polytechnique, RD128, F-91127 Palaiseau Cedex, France}
\author{Marie-Christine Angonin}\affiliation{LNE-SYRTE, Observatoire de Paris, CNRS UMR8630,
UPMC\\61 avenue de l'Observatoire, 75014 Paris, France}
\author{Peter Wolf}\affiliation{LNE-SYRTE, Observatoire de Paris, CNRS UMR8630,
UPMC\\61 avenue de l'Observatoire, 75014 Paris, France}

\date{\today}

\begin{abstract}
The efficiency of an atomic interferometer in proximity of a surface is discussed. We first study which is the best choice of frequency for a pulse acting on internal atomic transitions in the same well. Then considering the modification of atomic energy levels in vicinity of the surface, we propose the application of two simultaneous Raman lasers and numerically study the associated interference fringes. We show that the efficiency of the interferometric scheme is limited by the existence of a residual phase depending on the atomic path. We propose a symmetric scheme in order to avoid these contributions. We finally show that the suggested modifications make the contrast of the interference fringes close to 1 in any configuration, both close and far from the surface and with one or more initially populated wells.
\end{abstract}


\maketitle

\section{Introduction}

Atomic interferometry is an extremely powerful technique whose domains of application range from atomic clocks to gravimetry and fundamental physics. Its remarkable efficiency comes from the high precision that can be reached in the measurement of frequencies. Among its different applications, this experimental method allows to address the measurement of short-range interactions between an atomic ensemble and a macroscopic surface. Several experiments have been proposed and are currently being performed in this context \cite{LemondePRA05,WolfPRA07,DereviankoPRL09,SorrentinoPRA09}.

Due to the physics behind atomic interferometry, an accurate knowledge of the energy levels of the system is a fundamental issue as far as the optimization of the interferometric scheme is concerned. As we have shown in a recent paper \cite{MessinaPRA11}, these energy levels can be strongly modified by the proximity of a surface. This arises from the intrinsic modification of atomic wavefunctions due to the presence of a boundary condition and to the interaction between the atoms and the quantized electromagnetic field, known as Casimir-Polder interaction \cite{CasimirPhysRev48,ScheelActaPhysSlov08}, in presence of the surface.

Our calculations can be of interest for any interferometric measurement performed using an optical lattice in proximity of a surface. For the sake of clarity, we will focus our attention on the experiment FORCA-G (FORce de CAsimir et Gravitation \`a courte distance), recently proposed in order to measure the short-scale interactions between an atom of Rubidium and a massive surface \cite{LemondePRA05,WolfPRA07}. This experiment has two main goals. On one hand, it aims at providing a new measurement of the Casimir-Polder interaction. On the other hand, it also intends to set new constraints on the existence of a hypothetical deviation from the Newtonian law of gravitation predicted by unification theories \cite{OnofrioNewJPhys06}.

The experimental setup is based on a vertical standing wave producing an optical lattice. In an ideal periodic optical lattice, the additional linear gravitational potential leads to a class of localized atomic states in the trap, kwown as Wannier-Stark states \cite{Ashcroft,GluckPhysRep02}. As anticipated, these states and their energy levels are modified close to the surface \cite{MessinaPRA11}. An interferometric scheme is created by the way of a series of Raman pulses. This interferometric scheme is designed to separate the original wavepacket in two and to submit the two parts to a different potential by means of a spatial separation. After recombination, interferences are observed due to the phase shift cumulated during the time of separation in the trap. This phase shift depends on the potential difference supported by the two wavepackets at different distances from the surface. The measurement of the phase shift then gives access to the atom-surface interaction \cite{WolfPRA07}.
The main advantages of this experiment are the refined control of the atomic position as well as the high precision of the interferometric measurement. To exploit maximally these advantages, we have to optimize the interferometric scheme and the Raman pulses used to create it so that the contrast of the fringes at the end of the interferometer is maximal. In order to reach this goal, we can act on two parameters: the former is the frequency and intensity of the pulses used to move the atoms, the latter is the interferometric scheme itself. The main purpose of this paper is the optimization of these two parameters taking into account the real shape of the atomic wavefunctions in the trap as well as the modified energy levels. Nevertheless, we are going to show that this knowledge is not sufficient to design an efficient interferometric scheme, mainly because the presence of the surface breaks the translational symmetry characterizing the atomic states in an infinite optical lattice.

This paper is organized as follows. In Sec. \ref{Sec:2}, we present the physical system including the hamiltonian description of the atoms in the trap and the interaction of these atoms with the pulses used to create the interferometer. Section \ref{Sec:3} is dedicated to the optimization of the interferometric scheme initially suggested in \cite{WolfPRA07}. This section will be separated in two parts: on one hand we will present the optimization of the Raman pulses in proximity of the surface to take into account the modification of the atomic levels. On the other hand, we will discuss the problem arising from the scheme itself and we will propose a new interferometric scheme to solve it. Finally, the Sec. \ref{Sec:4} concern the results of this new scheme.

\section{The physical system}\label{Sec:2}

In this section, we are going to describe our physical system and the hamiltonian formalism used to investigate the dynamics of atoms trapped in front of a surface and submitted to a given series of laser pulses. Let us consider a two-level atom trapped in a vertical optical standing wave in proximity of a planar surface as described in \cite{MessinaPRA11}. The Hamiltonian of such a system is given by (with the same notations as in \cite{MessinaPRA11})
\begin{equation}\begin{split}
H&=H_0+H_\text{int}=H_\text{f}+H_\text{at}+H_{\footnotesize\text{WS}}+H_\text{int}\\
H_\text{f}&=\sumpk\,\hbar\omega\,\acpk\apk\\
H_\text{at}&=\hbar\omega_0\ket{e}\bra{e}\\
H_{\footnotesize\text{WS}}&=\frac{p^2}{2m}-mgz+\frac{U}{2}\bigl(1-\cos(2k_lz)\bigr)\\
H_\text{int}&=-\bbm[\mu]\cdot\mathbf{\mathcal{E}}(\mathbf{r}).\end{split}\label{Htot}\end{equation}
The complete Hamiltonian is written as a sum of a term $H_0$ describing the free evolution of the atomic and field degrees of freedom. In particular, $H_\text{f}$ is the Hamiltonian of the quantum electromagnetic field, described by a set of modes $(p,\mathbf{k},k_z)$: here $p$ is the polarization index, taking the values $p=1,2$ corresponding to TE and TM polarization respectively, while $\mathbf{k}$ and $k_z$ are the transverse and longitudinal components of the wavevector. $H_\text{at}$ is the internal Hamiltonian of our two level atom having ground state $\ket{g}$ and excited state $\ket{e}$ separated by a transition frequency $\omega_0$. While $H_\text{at}$ is associated to the internal atomic degrees of freedom, the term $H_{\footnotesize\text{WS}}$ accounts for the external atomic dynamics. As a consequence, it contains the kinetic energy ($p$ being the canonical momentum associated to $z$), as well as both the gravitational potential (treated here in first approximation as a linear term), where $m$
is the atomic mass and $g$ is the acceleration of the Earth's gravity, and the classical description of the stationary optical trap, having depth $U$. The interaction between the atom and the quantum electromagnetic field is written here in the well-known multipolar coupling in dipole approximation \cite{PowerPhilTransRoySocA59}, where $\bbm[\mu]=q\bbm[\rho]$ ($q$ being the electron's charge and $\bbm[\rho]$ the internal atomic coordinate) is the quantum operator associated to the atomic electric dipole moment and the electric field is calculated in the atomic position $\rv$.

This Hamiltonian can be separated in a part $H_0$ describing the free evolution of the atom (having ground and excited states $\ket{g}$ and $\ket{e}$ respectively, position $z$ and conjugate momentum $p$) and the quantum electromagnetic field (quantized in presence of a perfectly conducting surface in $z=0$ \cite{BartonJPhysB74}) and a perturbative term representing the interaction between the atom (having electric dipole moment $\bbm[\mu]$) and the field. As discussed in detail in \cite{MessinaPRA11}, the resolution of the time-independent Schr\"odinger equation for $H_0$ leads to a class of states similar to the well-known Wannier-Stark states \cite{GluckPhysRep02}, with a modification due to the presence of the surface. These states, noted with $\varphi_m(z)$ (where $m=1,2,\dots$), are identified with an index $m$ corresponding to the well of the standing wave in which the atom is trapped. Some examples of these states compared with the corresponding Wannier-Stark ones are provided in Fig.\ref{Psi}.
\begin{widetext}\begin{center}\begin{figure}[h!]\centering
\includegraphics[height=5cm]{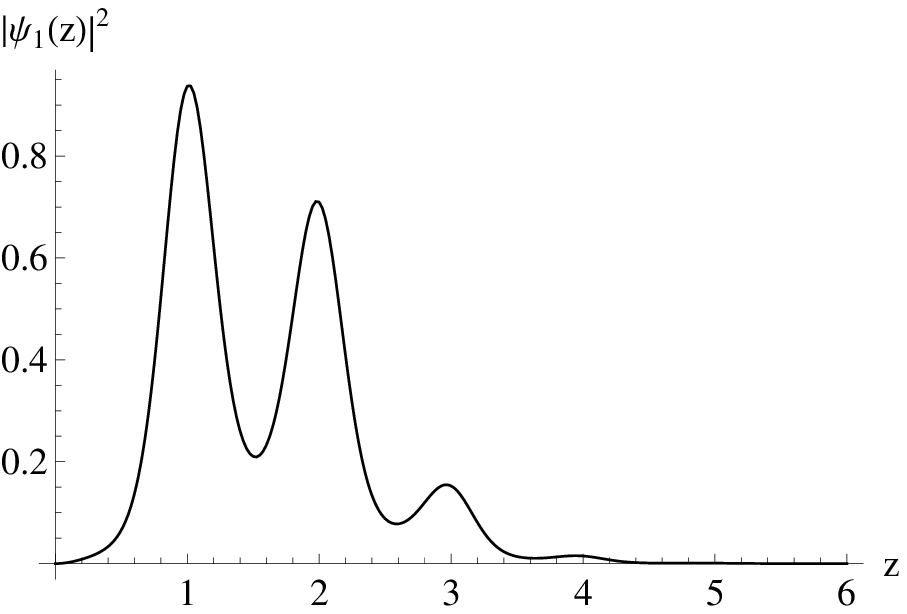}\hspace{1cm}\includegraphics[height=5cm]{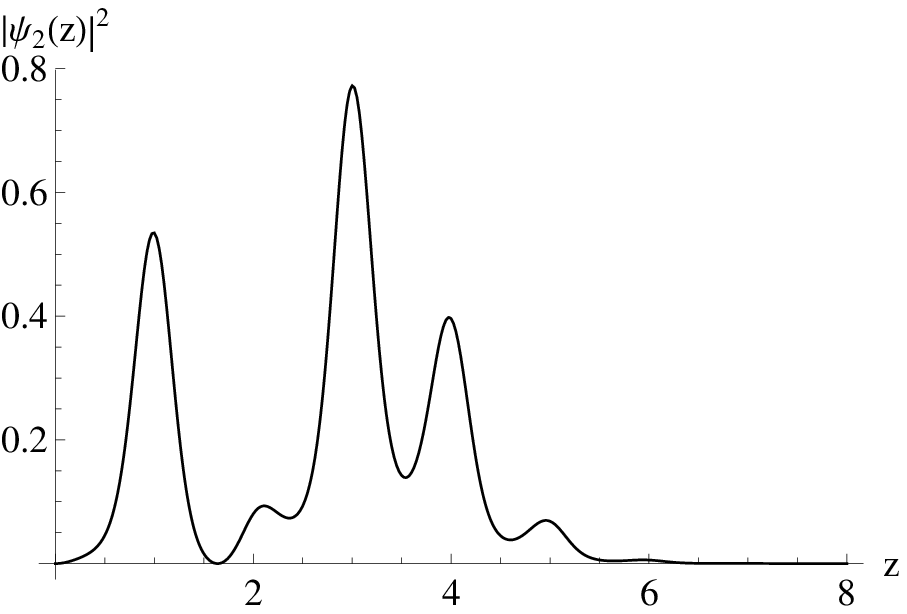}\\
\includegraphics[height=5cm]{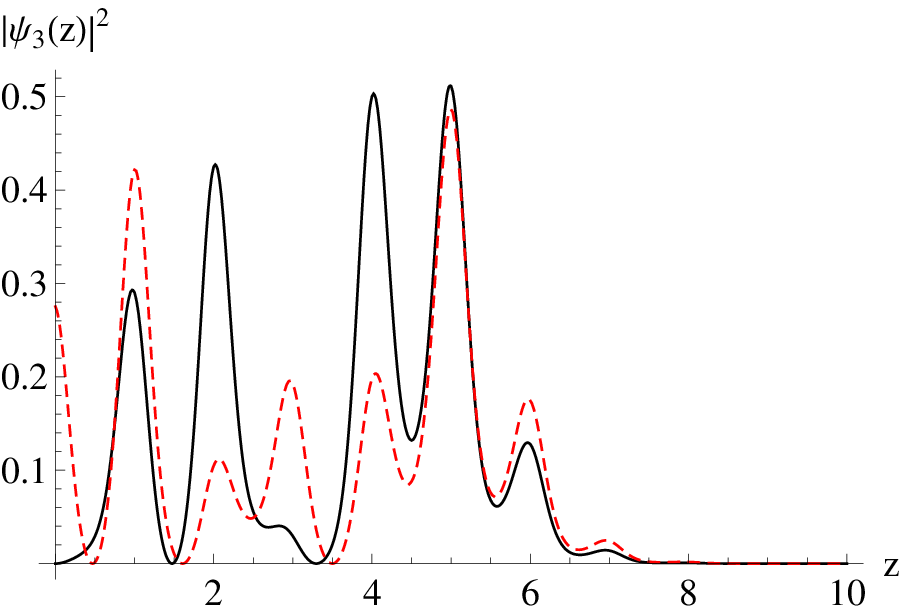}\hspace{1cm}\includegraphics[height=5cm]{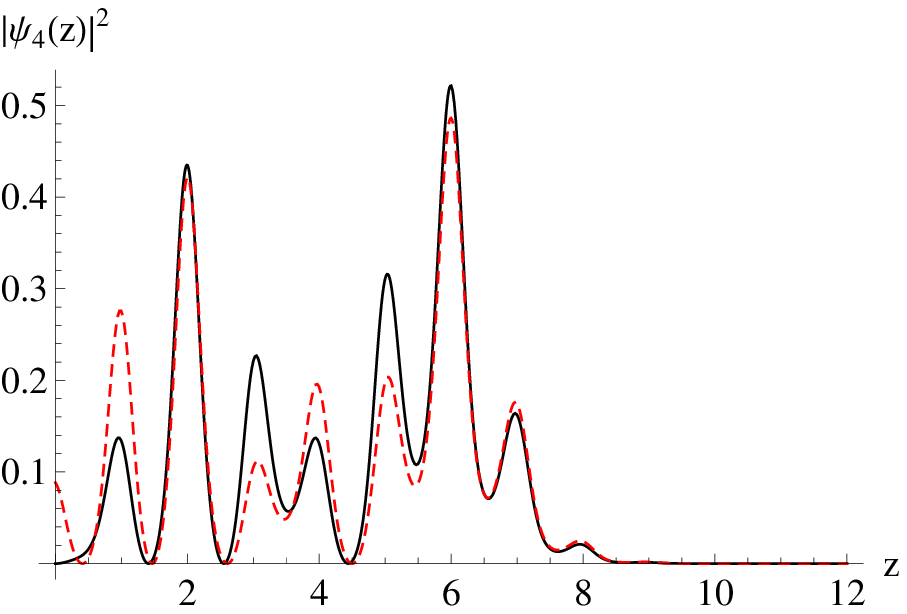}
\caption{(Color online) Square modulus of the first 4 modified Wannier-Stark states $\varphi_{m,1}(z)$ for $m=1,2,3,4$ and $U=3E_r$ where $E_r=\frac{\hbar^2k_l^2}{2m_a}$ represents the recoil energy of a photon of the laser \cite{MessinaPRA11}. The last two functions (black, solid line) are compared to the corresponding standard Wannier-Stark state (red, dashed line). Here, the position $z$ of the atom is expressed as a function of the periodicity of the trap $\frac{\lambda_l}{2}$.}\label{Psi}\end{figure}\end{center}\end{widetext}

\begin{figure}[h]\centering
\includegraphics[height=5cm]{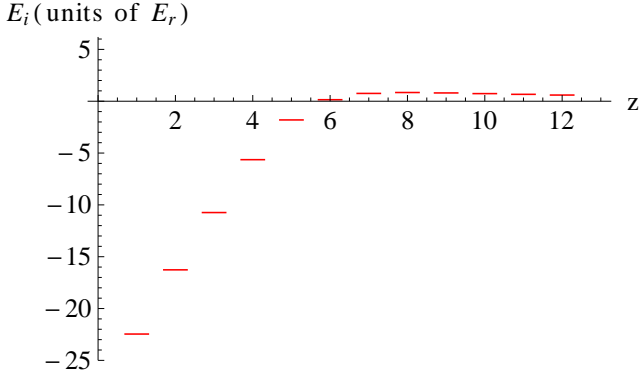}
\caption{Value of the first twelve energy levels in presence of the surface and taking into account the Casimir-Polder effect. Once again $z$ is expressed in units of $\frac{\lambda_l}{2}$.}\label{nivE}\end{figure}

Moreover, the perturbative treatment of the atom-field interaction term leads to a shift of the energy levels of $H_0$. This shift is the result of the well-known Casimir-Polder effect \cite{ScheelActaPhysSlov08} and the modified energy levels are shown in Fig.\ref{nivE}.

\begin{figure}\hspace{-0.5cm}
\includegraphics[height=5cm]{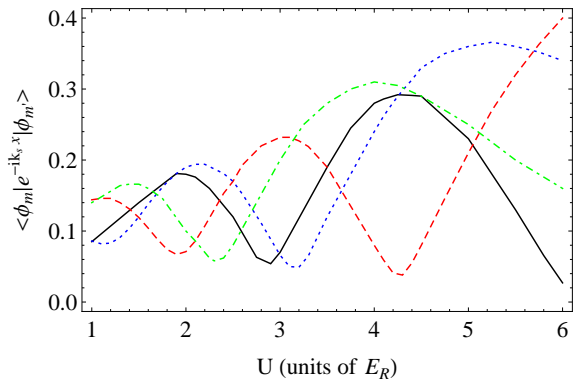}
\caption{(Color online) Normalized Rabi frequencies for $m-m'=0,\pm1,\pm2,\pm3$ transitions, as a function of lattice depth. The Black solid line represents the Rabi frequency for $m-m'=0$, the red dashed line is the Rabi frequency for $m-m'=\pm1$, the blue dotted line is for $m-m'=\pm2$ and the green dot-dashed line stands for $m-m'=\pm3$. Here, the wavelength chosen for the trap is $\lambda_l=532\,$nm which is the case of FORCA-G.}
\label{rabi}
\end{figure}

In addition to the stationary Hamiltonian \eqref{Htot}, we have to take into account the presence of laser pulses tuned on the atomic transition energy plus the energy difference between two different wells.
The potential representing the atom-laser interaction can be written as
\begin{equation}H_s=\hbar\,\Omega\,\cos(\omega_st-k_sx)\ket{e}\bra{g}+\text{H.c.}\end{equation}
where $\Omega$ is the Rabi frequency while $\omega_s$ is the probe laser frequency and $k_s$ its wavevector.
In order to model the interferometer scheme and to adjust the frequency of the beams to maximize the contrast of matter-wave interferences, we have to solve the time-dependent Schr\"odinger equation
\begin{equation}\label{DiffEq}i\hbar\frac{d}{dt}\ket{\Psi_\text{at}(t)}=(H+H_s)\ket{\Psi_\text{at}}.\end{equation}
The resolution of Eq. \eqref{DiffEq} will be performed by projecting it on the basis of external and internal atomic states $\{\ket{\varphi_m,i}=\ket{\varphi_m}\otimes\ket{i}\}$ where $m=1,2,\dots$ and $i=e,g$. In this basis, the atomic wavefunction can be expressed as
\begin{equation}\begin{split}
\ket{\Psi_\text{at}(t)}=&\sum_m\Bigl[a_m^g(t)e^{-i\frac{E_mt}{\hbar}}\ket{\varphi_m,g}\\&\,+a_m^e(t)e^{-i\frac{E_m+\hbar\omega_{eg}}{\hbar}t}\ket{\varphi_m,e}\Bigr].
\end{split}\end{equation}
Using this expression, the Schr\"odinger equation is equivalent to the following set of coupled equations for the coefficients $a_m^g(t)$ and $a_m^e(t)$
\begin{equation}\begin{split}
i\dot{a}_m^g(t)&=\sum_{m'}\Omega\,\bra{\varphi_m}\cos(\omega t-k_sx)\ket{\varphi_{m'}}a_{m'}^e(t)e^{-i\delta_{m'-m}t}\\
i\dot{a}_m^e(t)&=\sum_{m'}\Omega\,\bra{\varphi_m}\cos(\omega t-k_sx)\ket{\varphi_{m'}}a_{m'}^g(t)e^{i\delta_{m-m'}t}
\end{split}\end{equation}
where $m'$ runs over all the wells and we have defined $\delta_{m-m'}=\frac{E_{m}-E_{m'}+\hbar\omega_{eg}}{\hbar}$. Making the usual rotating wave approximation \cite{LemondePRA05} and defining
\begin{equation}
\Delta_{m,m'}=\omega-\delta_{m-m'}=\omega-\omega_{eg}-\frac{E_{m}-E_{m'}}{\hbar}
\end{equation}
 we finally get
\begin{equation}\begin{split}
i\dot{a}_m^g(t)&=\sum_{m'}\frac{\Omega}{2}\,a_{m'}^e(t)e^{i\Delta_{m',m}t}\bra{\varphi_m}e^{-ik_sx}\ket{\varphi_{m'}}\\
i\dot{a}_m^e(t)&=\sum_{m'}\frac{\Omega}{2}\,a_{m'}^g(t)e^{-i\Delta_{m,m'}t}\bra{\varphi_m}e^{ik_sx}\ket{\varphi_{m'}}.
\label{coeff}\end{split}\end{equation}
In Eq. \eqref{coeff}, the term $\frac{\Omega}{2}\,\bra{\varphi_m}e^{-ik_sx}\ket{\varphi_{m'}}$ is the Rabi frequency for the transition between two different wells $m$ and $m'$, $\Omega$ being the Rabi frequency in free space \cite{BeaufilsPRL11}. This frequency governs the transition probability from one well to another and it strongly depends on the depth of the trapping potential and on the
trapping wavelength. In Fig. \ref{rabi} we show the behavior of these frequencies as a function of the well depth for our chosen trapping wavelength.
The choice of these frequencies is fundamental for the realization of our interferometer scheme and it will determine the experimental parameters. In particular, the lattice depth will be chosen as $U=3E_r$ in order to maximize the probability for the transition $m-m'=\pm1$ and minimize the probability for $m-m'=\pm2$ and $m-m'=0$ \cite{TackmannPRA11}. In the following, we will study the optimization of the interferometric scheme with this depth.

\section{Optimization of the interferometric scheme}\label{Sec:3}

The choice of the interferometric scheme is fundamental in order to have a precise measurement of the short-range interactions we intend to investigate. The basic idea of atomic interferometry is to create a coherent superposition of the two internal states in the starting well, to move the two components in the two neighbouring wells and to finally recombine them in the starting well. After recombination, the different path followed by the two components of the wavepacket leads to interference fringes due to the phase shift which is related to the energy difference between the two external wells. This section is dedicated to the investigation of an optimal interferometric scheme for our purpose. In the first section, we will discuss the problem of the interaction time between the Raman pulses and the atoms to create appropriate superposition. Then we will study the first proposal of interferometric scheme for FORCA-G \cite{LemondePRA05,WolfPRA07} when the atoms are close to the surface so that their energy levels are modified by the Casimir-Polder interaction and a hypothetical Yukawa gravitational deviation. Finally we will investigate the problem arising when several wells are populated at the beginning of the interferometer and we will propose a new scheme for the experiment.

\subsection{Optimization of the sequence of pulses}

\begin{figure}[h]\centering
\includegraphics[height=4.5cm]{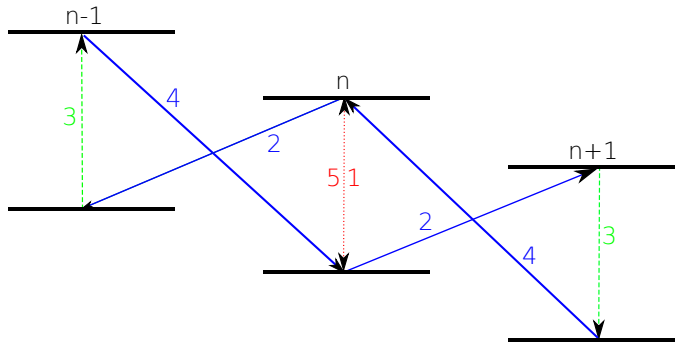}
\caption{(Colors online) Butterfly scheme for the interferometer \cite{WolfPRA07} where the red dotted line represents the microwave $\frac{\pi}{2}$-pulses used to create the coherent superposition and for the recombination at the end of interferometer. The blue solid lines represent the Raman $\pi$-pulses toward neighbouring wells and the green dashed lines are the microwave $\pi$-pulses in the two separated wells.}
\label{butterfly}
\end{figure}

 In the first proposal presented in Fig. \ref{butterfly} of the experiment FORCA-G \cite{WolfPRA07}, the suggested interferometric scheme was a \emph{butterfly}-like scheme with a trap depth of $U=3E_r$.
As it can be seen in Fig. \ref{butterfly} the interferometer consists in a sequence of pulses aiming at separating the atomic wavepacket in two parts and to recombine them after a time $T$ of free evolution in two separated wells.
\newline

In order to realize this scheme, we need at least 5 Raman pulses to create the coherent superposition in the starting well, to move it in neighbouring wells and to recombine it in the starting well. We can remark here that it is in principle possible to add pulses in order to reach farther wells and then obtain a bigger phase shift in the interference pattern. However, the trapping time of the atoms is limited by experimental conditions and it is typically of the order of 1\,s. At the same time, the free-evolution time has to be as long as possible in order to maximize the number of periods in interference fringes. As a consequence, it is crucial to optimize the sequence of pulses in order to make them as short as possible.

We have seen in Sec. \ref{Sec:2} that the depth of the trap was chosen to favour the transition between adjacent wells. As a consequence, the transition between the internal atomic states in the same well (corresponding to the pulses 1, 3 and 5 on Fig. \ref{butterfly}) is not efficient.

A natural solution to solve this problem is to use a microwave pulse instead of a Raman laser for the atomic transitions in the same well. Indeed, the coupling between the external states of the atom and the laser in Eqs. \eqref{coeff} can be expressed as
\begin{equation}\bra{\varphi_m}e^{-ik_sx}\ket{\varphi_{m'}}=\int_{-\infty}^{+\infty}\varphi_{m}^*(x)\varphi_{m'}(x)e^{-ik_sx}\,dx,\label{coupling}\end{equation}
and whereas this coupling depends on the shape of the wavefunctions in the case of Raman beams, this is much less the case for microwave pulses. In fact, for these pulses the wavevector is such that $k_sx\ll1$, so the exponential term in Eq. \eqref{coupling} tends to 1 and Eq. \eqref{coupling} is reduced to the scalar product $\bra{\varphi_m}\varphi_{m'}\rangle=\delta_{m,m'}$. As a consequence, the microwave pulses are much more efficient for transitions in the same well and the duration of such pulses is very short in comparison with the duration of Raman pulses.

\subsection{The interferometer close to the surface}
Once the sequence of Raman and microwave pulses is chosen, we can investigate the interferometric scheme presented in Fig. \ref{butterfly} close to the surface where the atomic energy levels are modified.
In this scheme, the atoms are first in the ground state $\ket{g}$. We apply a first $\frac{\pi}{2}$-microwave pulse tuned on the atomic transition in order to create a coherent superposition of the two internal states $\ket{g}$ and $\ket{e}$. We then apply a $\pi$-laser pulse with a frequency tuned on the atomic transition plus a detuning corresponding to the energy difference with the neighbouring wells in order to move the two components of the atomic wavefunction in two different wells. We let the system evolve in the arrival state during a time $T$ before inducing a transition in the two separated wells using a $\pi$-microwave pulse. After an additional free-evolution time $T$ in the separated wells, we finally induce the inverse path and recombine the wavefunction to observe interference fringes due to the energy difference between the two wells. To simulate this interferometer, Eqs. \eqref{coeff} are solved numerically using a C++ program.
At the starting point of the simulation, we consider the atom in the ground atomic state $\ket{g}$ in the well $m$. We thus solve the coupled Eqs. \eqref{coeff} with the initial conditions
\begin{equation}
\left\{
\begin{array}{ll}
a_n^g(t=0)&=0 \qquad \forall n\neq m\\
a_m^g(t=0)&=1\\
a_n^e(t=0)&=0 \qquad \forall n
\label{CI}
\end{array}
\right.
\end{equation}
and with a microwave laser pulse of frequency $\nu_{\text{mw}}=6.8\,$GHz and $\Omega=100\,$rad/s with a duration of $t_1=0.0157\,$s, corresponding to a $\frac{\pi}{2}$-pulse. Once the coherent superposition is created we apply the series of pulses shown in Fig. \ref{butterfly}, taking as initial condition the end of the previous pulse. As it can be seen in Fig. \ref{nivE}, when the atom is close to the surface, its energy levels are modified mainly by the Casimir-Polder effect so that the energy differences between a given well and its two adjacent wells are no longer equal as in the absence of the surface \cite{Ashcroft}. In order to study the effect of this difference, we first used one single laser for each Raman pulse taking for its frequency the mean of the two energy differences. However, with this choice, we have numerically verified that the contrast at the end of the interferometer remains close to zero when the initially populated well is close to the surface. This is not surprising, since in this case the Raman lasers are too detuned with respect to each transition frequency. We have verified this for atoms initially in the well 8. In this well, the energy differences playing a role in the interferometer is the differences with the wells 7 and 9. These differences are respectively
\begin{equation}\begin{split}
\Delta E_{7,8}&\simeq0.7013 E_r\\
\Delta E_{8,9}&\simeq0.0393 E_r.
\label{diff8}\end{split}\end{equation}
As a consequence, when we use a Raman pulse with $\Omega=100\,$rad/s ($\hbar\Omega=1.23.10^{-2}E_R$) this pulse is relatively thin in the frequency domain with respect to the difference in energies \eqref{diff8} and thus the states are not efficiently poulated. One could increase $\Omega$, but in that case the pulses would be too broad and significantly populate additional wells (e.g. well 7), which again leads to a loss of contrast.

\begin{figure}[h!]\centering
\includegraphics[height=5cm]{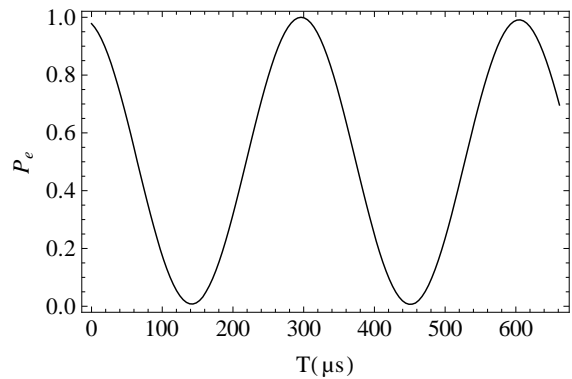}
\caption{Interference fringes at the end of the butterfly interferometer as a function of the free-evolution time in the case of two lasers used for the Raman transitions. The starting well is $m=8$.}
\label{twolasers}
\end{figure}

A natural solution to improve the contrast of the interferometer is to take two Raman lasers with two different frequencies, each tuned on one energy difference. The interference pattern in this configuration is shown in Fig. \ref{twolasers} for the starting well 8.
We can see that the contrast is, in this configuration, close to 1.

\subsection{The interferometer in the case of several initially populated wells}

\begin{figure}
\includegraphics[height=5cm]{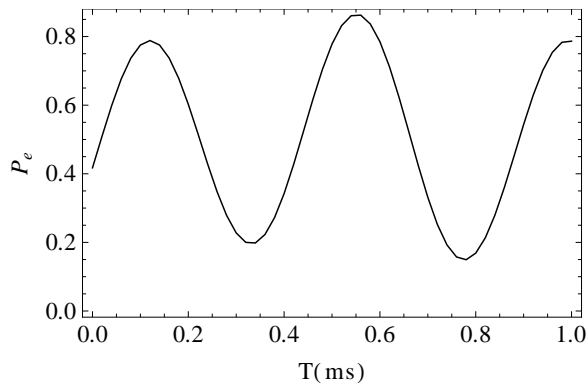}
\caption{Interference fringes at the end of the butterfly interferometer in function of the free-evolution time in the case of two initially populated wells. Here are shown two wells far from the surface in a regime of standard Wannier-Stark states.}
\label{twowells}
\end{figure}

We have seen in the previous section that using two Raman lasers for the transitions toward neighbouring wells makes the contrast of the interferometer in the case of one initially filled well maximal even for atoms very close to the surface. Nevertheless it is experimentally difficult to populate one single well at the beginning of the measurement. At the same time, an initial condition with more than one populated well far from the surfcae could be of interest in order to increase the signal-noise ratio and thus the precision in the measurement of a hypothetical Yukawa deviation far from the surface and of the Earth gravitational acceleration $g$, which constitutes an important test for the experiment. The resulting contrast in the case of two initially populated wells far from the surface (in the Wannier-Stark regime) is shown in Fig. \ref{twowells}.
We can see that the contrast is relatively low even in the case of wells taken far from the surface in a region where the energy separation between two adjacent wells is well-independent. The explanation of this phenomenon lies in the properties of the standard Wannier-Stark states. As a matter of fact, these states present a translational symmetry which allows us to rewrite Eqs. \eqref{coeff} as a function of the state in the central well labelled $0$ (being the Wannier-Stark states defined in $\left[-\infty,\infty\right]$) \cite{LemondePRA05}.
\begin{equation}\begin{split}
i\dot{a}_m^g(t)&=\sum_{m'}\Omega_{m-m'}^*\,a_{m'}^e(t)e^{i\Delta_{m',m}t}e^{-i\pi m'\frac{k_s}{k_l}}\\
i\dot{a}_m^e(t)&=\sum_{m'}\Omega_{m'-m}\,a_{m'}^g(t)e^{-i\Delta_{m,m'}t}e^{i\pi m\frac{k_s}{k_l}}
\label{sym}
\end{split}\end{equation}
with $\Omega_{m}=\frac{\Omega}{2}\bra{\text{WS}_0}e^{ik_sx}\ket{\text{WS}_m}$.
In Eqs. \eqref{sym}, we can see that expressing the coupling elements as a function of the central well introduces a phase factor of the form $e^{i\pi m'\frac{k_s}{k_l}}$ with a different sign depending on the transition. There is a minus when we go from ground to excited state and conversely so that the phase difference at the end of the interferometer can be expressed as
\begin{equation}\begin{split}
\Delta\phi=&-4m\pi\frac{k_s}{k_l}+\frac{2}{\hbar}\left(m_ag\lambda_l+U_{m+1}-U_{m-1}\right)T\\
&-\phi_s^{(1)}+2(\phi_s^{(2)}-\phi_s^{(3)}+\phi_s^{(4)})-\phi_s^{(5)}.
\label{dephas}
\end{split}\end{equation}
In Eq. \eqref{dephas}, the terms $\phi_s^{(i)}$ stand for the phase of each pulse and can be evaluated whereas the term of interest is the term $\frac{2}{\hbar}\left(m_ag\lambda_l+U_{m+1}-U_{m-1}\right)T$ where $U_m$ is the external potential (Casimir and Yukawa) in the well $m$.

As a consequence, for two initial wells the phase term $-4m\pi\frac{k_s}{k_l}$ is different because of the factor $m$ which is the starting-well index. So, at the end of the interferometer, we obtain two interference contributions with different phases which drastically reduce the contrast. This effect is even worse when more than two wells are initially populated. So we have to find a new interferometric scheme which could avoid these phase shifts, i.e. where the Raman pulses are symmetric in order to get rid of the contribution proportional to $m$. We will describe the new scheme in the next section.

\section{Results for the modified interferometric scheme}\label{Sec:4}
\begin{figure}[h!]\centering
\includegraphics[height=4.5cm]{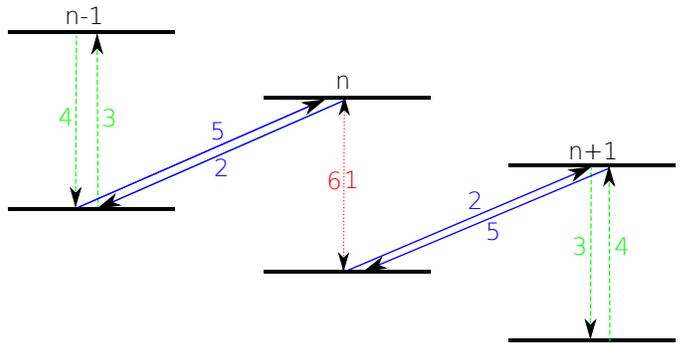}
\caption{(Colors online) New symmetric interferometric scheme where the red dotted line represents the microwave $\frac{\pi}{2}$-pulses used to create the coherent superposition and for its recombination at the end of interferometer. The blue solid lines represents the Raman $\pi$-pulses in neighbouring wells and the green dashed lines are the microwave $\pi$-pulses in the two separated wells.}
\label{interfero}
\end{figure}

As we have seen in the previous section, the butterfly scheme for the interferometer does not maximize the contrast of the interferences fringes. To avoid the problem of the phase shift cumulated during the Raman pulses, we propose a new scheme of interferometer with an additional microwave $\pi\,$-pulse in the wells $m\pm1$ as shown in Fig. \ref{interfero}.
This pulse aims at symmetrizing the scheme leading to a cancellation of the residual well-dependent phase present in the case of the butterfly interferometer (see e.g. Eq. \eqref{sym}).
As the atom follows the same path during separtation and recombination, there is no additional phase factor on each arm , and the total phase difference is independent of the starting well index.

However, adding a new pulse extends the duration of the interferometric scheme. So we have to optimize the length of the pulses as we have done in the case of the butterfly scheme. Table \ref{pulses} shows the optimal duration of each pulse, together with the final population of states $\ket{g}$ and $\ket{e}$ in the wells of interest (with $\Omega=100\,$rad/s). These interaction times are found by solving Eqs. \eqref{coeff} and optimizing the duration for each pulse. As in the case of the butterfly scheme, we use microwave pulses to induce transitions in the same well and Raman pulses for the transitions between adjacent wells. For the example presented in Table \ref{pulses}, we have considered the well number 2 as starting well.

The duration of the first pulse is then given by the first cross of populations in the starting well ($\frac{\pi}{2}$-pulse) and at the inversion of population for the pulses 2, 3, 4 and 5 ($\pi$-pulse).
Concerning the last pulse, it's duration is given by the maximal population of $\ket{e}$ ($\frac{\pi}{2}$-pulse). Each condition was found with a dichotomy algorithm in C++ \cite{Press95}.

\bigskip
\begin{center}\begin{table}\begin{tabular}{|c||c|c||c|}
\hline
Pulse & \multicolumn{2}{c||}{Duration (s)} & Population \\
\hline\hline
1 & \multicolumn{2}{c||}{$0.0157$} &  $P_g(2)=0.5000$ \\
 & \multicolumn{2}{c||}{} & $P_e(2)=0.5000$ \\
\hline\hline
2 & laser 1 & laser 2 & $P_g(1)=0.5002$ \\
\cline{2-3}
 & $0.1500$ & $0.0974$ & $P_e(3)=0.4983$ \\
 \hline\hline
3 & \multicolumn{2}{c||}{$0.0314$} &  $P_g(3)=0.4983$ \\
 &\multicolumn{2}{c||}{} & $P_e(1)=0.5002$ \\
\hline\hline
4 & \multicolumn{2}{c||}{$0.0314$} &  $P_g(1)=0.5002$ \\
 & \multicolumn{2}{c||}{} &  $P_e(3)=0.4983$ \\
\hline\hline
5 &  laser 1 & laser 2 & $P_g(2)=0.5001$ \\
\cline{2-3}
 & $0.1618$ & $0.0974$ & $P_e(2)=0.5004$ \\
 \hline\hline
6 & \multicolumn{2}{c||}{$0.0157$} &  $P_g(2)=0$ \\
 & \multicolumn{2}{c||}{} & $P_e(2)=1.001$ \\
\hline
\end{tabular}
\caption{Duration and population in the wells of interest after each pulse. In this table, the number in brackets represent the well under scrutiny.}
\label{pulses}
\end{table}\end{center}

\begin{center}
\begin{figure}[h!]
\includegraphics[height=5cm]{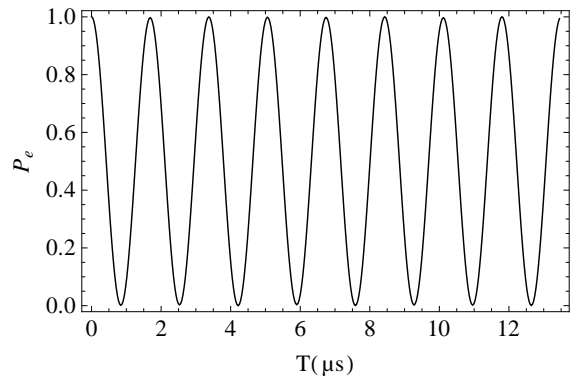}
\caption{Contrast of the interferences fringes at the end of the symmetric interferometer as a function of the free-evolution time $T$. As an example, we show here the contrast for all atoms initially in well $m=2$.}
\label{contrast1}
\end{figure}\end{center}

\begin{center}
\begin{figure}[h!]
\includegraphics[height=5cm]{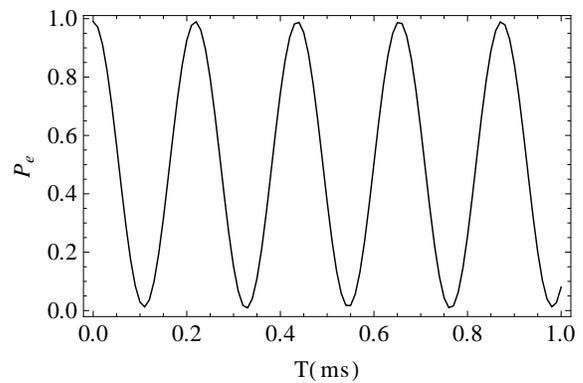}
\caption{Contrast for an initial superposition of two wells very far from the surface (Wannier-Stark regime).}
\label{contrast2}
\end{figure}\end{center}

After optimization of the pulses, the total duration of our interferometer, without counting the free evolution time, is around $0.4\,$s.
We now have to test this new scheme by analyzing the contrast at the end of the interferometer.
In Figs. \ref{contrast1} and \ref{contrast2} we plot the contrast obtained with optimized duration for the pulses as a function of the free-evolution time. We can see that the contrast is maximal even in the case of two wells populated at the beginning. Moreover, the typical oscillation period is much shorter than the coherence time of the atomic ensemble.

\section{Conclusions}

In this paper, we have discussed the optimization of the interferometric scheme for an atomic ensemble trapped in proximity of a surface.  In particular, we have started our discussion by studying the efficiency of the interferometric scheme initially proposed in \cite{LemondePRA05,WOLFPRA07} for the experiment FORCA-G.

As a first step, we have shown that in order to minimize the interaction time, it is more efficient to use microwave pulses instead of Raman pulses to induce atomic internal transitions in one single well.

We have then considered the effect of the surface: the atomic energy levels in its proximity are strongly modified (mainly by the Casimir-Polder interaction) destroying the translational symmetry of the energy differences. This drastrically reduces the contrast at the end of the interferometer. We have shown that this problem is solved by using two different simultaneous Raman lasers, each tuned on one energy difference.

Next we have seen that the butterfly scheme proposed in \cite{LemondePRA05,WolfPRA07} induces a loss of contrast due to the cumulation of residual phases during the Raman pulses in neighboring wells. A new scheme has been suggested to solve this problem. This consist in an addition of a symmetrization pulse in the two separating wells. As a consequence, the resulting scheme presents symmetric Raman pulses which lead to a cancellation of the residual phases. Finally, we have shown that the modified interferometric scheme we propose leads to a contrast near to 1 both in proximity and far from the surface, optimizing the conditions in which the experiment is performed.

Our analysis can be of interest for any interferometric experiment performed in proximity of a surface. However, it was demonstrated that the Wannier-Stark states have a finite lifetime in the trap due to Landau-Zener tunneling \cite{ZenerProcRSocLondA34,GluckEurJPhysD98}. This effect was not considered in this paper where we have assumed that this lifetime is very long in comparison with the trapping time of the atoms limited by experimental conditions. This is the case for the standard Wannier-Stark states for which the lifetime is of the order of $10^{15}\,$s (for FORCA-G parameters) whereas the experimental trapping time is around 1\,s. However, it remains to be understood how the presence of the surface influences this lifetime and this will be the main subject of an upcoming publication.

\begin{acknowledgments}
This research is carried on within the project iSense, which
acknowledges the financial support of the Future and Emerging
Technologies (FET) programme within the Seventh Framework
Programme for Research of the European Commission, under FET-Open
grant number: 250072. We also gratefully acknowledge support by
Ville de Paris (Emergence(s) program) and IFRAF. The authors thank
Q. Beaufils, G. Tackmann, A. Hilico, B. Pelle and F. Pereira dos Santos for fruitful and
stimulating discussions.
\end{acknowledgments}

\end{document}